\documentclass[twocolumn,trackchanges,times]{aastex701}

\usepackage{standalone}
\standaloneconfig{mode=buildnew}

\usepackage{amsmath}
\allowdisplaybreaks[4]

\begin{document}

\title{Detecting Exoplanets beyond Local Supercluster through Gravitational Waves with B-DECIGO and DECIGO}

\author[0009-0000-9276-5665]{Wen-Long Guo}
\affiliation{School of Physics and Astronomy, Beijing Normal University, Beijing 100875, China}
\email{guowl@mail.bnu.edu.cn}

\author[0000-0003-3328-9448]{Li-Ming Zheng}
\affiliation{School of Physics and Astronomy, Beijing Normal University, Beijing 100875, China}
\affiliation{School of Physics and Astronomy, Cardiff University, Cardiff, CF24 3AA, United Kingdom}
\email[show]{lmzheng@mail.bnu.edu.cn}

\author[0000-0002-8492-4408]{Zhengxiang Li}
\affiliation{School of Physics and Astronomy, Beijing Normal University, Beijing 100875, China}
\affiliation{Institute for Frontiers in Astronomy and Astrophysics, Beijing Normal University, Beijing 102206, China}
\email[show]{zxli918@bnu.edu.cn}

\author[0000-0002-3567-6743]{Zong-Hong Zhu}
\affiliation{School of Physics and Astronomy, Beijing Normal University, Beijing 100875, China}
\affiliation{Institute for Frontiers in Astronomy and Astrophysics, Beijing Normal University, Beijing 102206, China}
\affiliation{Department of Astronomy, School of Physics and Technology, Wuhan University, Wuhan 430072, China}
\email{zhuzh@bnu.edu.cn}

\begin{abstract}
    The first detection of a gravitational-wave (GW) signal in 2015 has opened a new observational window to probe the universe. This probe can not only reveal previously inaccessible binaries, black holes, and other compact objects, but also can detect exoplanets through their imprint on GW signals, thereby significantly extend current exoplanet surveys. To date, nearly 6000 exoplanets have been confirmed, yet most of them reside either in the solar neighbourhood or along the sightline toward the Galactic bulge, reflecting the range limits of traditional electromagnetic techniques. In this work, we adopt the method proposed in \citet{tamaniniGravitationalwaveDetectionExoplanets2019} to investigate frequency modulations in GW signals from early-stage binary neutron stars (BNSs) induced by circumbinary planets (CBPs) and find that CBPs can be detected by the future space-borne detector Deci-hertz Interferometer Gravitational-wave Observatory (DECIGO). For BNS system with the masses of two components both being 1.4 $M_{\odot}$, DECIGO could detect CBPs with mass being dozens of times that of Jupiter out to distances of $\sim 1$ Gpc, well beyond the Local Supercluster, offering an unprecedented opportunity to study planetary formation and evolution for the post-main-sequence stage.
\end{abstract}


\section{Introduction}\label{sec:intro}
    Over the past two decades, observations in electromagnetic bands have delivered breakthrough advances in exoplanet science, revealing a coherent continuum from protoplanetary disks to mature planetary systems. Owing to the extraordinary high angular resolution observations of the Atacama Large Millimeter Array (ALMA), the early processes of planet formation have been systematically revealed through structures~\citep{huangDiskSubstructuresHigh2018}, gas kinematic evidence~\citep{pinteKinematicEvidenceEmbedded2018}, chemical environment~\citep{obergMoleculesALMAPlanetforming2021}, and statistical properties~\citep{ansdellALMASurveyLupus2016}, of protoplanetary disks. Meanwhile, a large number of stable planetary systems around main sequence host stars have been accumulated via different methods, i.e. transit \citep{rickerTransitingExoplanetSurvey2015,thompsonPlanetaryCandidatesObserved2018}, radial velocity \citep{mayorHARPSSearchSouthern2011,pepeHARPSSearchEarthlike2011}, astrometry \citep{hollGaiaDR3Astrometric2023}, and gravitational microlensing observations \citep{cassanOneMoreBound2012,suzukiExoplanetMassratioFunction2016}. To date, nearly 6000 exoplanets have been confirmed~\footnote{https://exoplanetarchive.ipac.caltech.edu}. However, in terms of the space distribution, most of these protoplanetary disks and mature planet-star systems reside only either in the solar neighborhood or along the sightline toward the Galactic bulge. In terms of the evolution stage, most of these systems are only in their infantile and postadolescent stages. That is, the vast majority of their host stars are in the proto or main sequence phases~\citep{christiansenNASAExoplanetArchive2025}. As a result, the above-mentioned incompleteness would lead to pronounced observational selection effects and environmental bias in our understanding for the formation and evolution of planet-star systems.
    
    Moreover, the survival and possible re-formation of planets in the post–main-sequence phase—particularly after violent conditions or explosive events such as common-envelope (CE) episodes and supernovae—remain poorly understood. 
    %Compact object systems offer a unique probe for planetary science under extreme conditions. 
    Detections of pulsar planets \citep{wolszczanPlanetarySystemMillisecond1992, wolszczanConfirmationEarthmassPlanets1994} and even one special case, the confirmed circumbinary planets (CBPs) in the neutron star-white dwaft binary system PSR B1620--26 \citep{thorsettPSRB162026Binary1993, sigurdssonYoungWhiteDwarf2003} suggest that the survival of the first-generation planets and the formation of the second-generation planets may be feasible. However, for compact object binary systems that have experienced violent processes or CE phases, questions about whether planets can form and be long-lived, the relevant formation windows and material reservoirs, and the upper limits on CE mass transfer, and dynamical constraints remain essentially untested by observations. Addressing these questions requires a detection channel that circumvents dust extinction and the limitations of electromagnetic visibility. 
    
    To overcome these limitations, researchers have explored GW approaches to exoplanet detection. However, even considering GW produced by eccentric systems at higher harmonics, or oscillation modes of the binaries, viable detectable systems were found at the beginning \citep{ferrariGravitationalWavesEmitted2000, bertiExcitationGmodesSolartype2001}. As space-based GW missions take shape, studies have proposed employing LISA to probe exoplanetary systems \citep{setoDetectingPlanetsCompact2008, wongPossibilityDetectingUltrashort2019}. Inferring the source’s radial velocity from the Doppler shifts imprinted on the GW phase by a planetary companion has been proposed, thereby identifying CBPs and assessing the implications for planetary formation and evolution theory \citep{tamaniniGravitationalwaveDetectionExoplanets2019,  danielskiCircumbinaryExoplanetsBrown2019, danielskiWillGravitationalWaves2020, kangProspectsDetectingExoplanets2021, katzBayesianCharacterisationCircumbinary2022}. However, analogous studies for binary neutron stars systems that typically undergo two collapses and likely experience a common-envelope phase remain sparse. Detecting such planetary systems would not only bridge the electromagnetic-sample gap in extreme environments but also provide direct observational samples to constrain the environmental limits of planet formation, the existence and occurrence of second-generation planets, and the quantitative bounds on common envelope mass transfer. In this work, we adopt and extend the method in \cite{tamaniniGravitationalwaveDetectionExoplanets2019} to assess the prospects to detect CBPs of both binary neutron star (BNS) and double white dwarfs (DWDs) with DECIGO and B-DECIGO. We also systematically reviewed the instrument response and sensitivity curves of space-based detector. Moreover, we discussed the detection limit and found it could push the detection reach from $\sim 1$ kpc scale, typical of electromagnetic surveys, and $\sim 1$ Mpc scale, limit of LISA \citep{danielskiWillGravitationalWaves2020}, to $\sim 1$ Gpc scale luminosity distances, increasing the accessible distance by six orders of magnitude over that of electromagnetic surveys, extending the exoplanet detection volume well beyond the Local Supercluster and even out to Abell galaxy clusters. Finally, we explored the dependence of detectability on  luminosity distance, planet masses, masses of source components, and the GW frequency of sources.
    
    This paper is organized as follows. In section~\ref{sec:meth}, we present a detailed description of the detection methodology and review the response of space-based gravitational wave detectors. In section~\ref{sec:result}, we present the results and discuss the detection limit and key determinants. Finally, we present the conclusions in section~\ref{sec:conclu}. Throughout this paper we use geometrized units with $G = c = 1$.

\section{Method}\label{sec:meth}

\subsection{Space-borne Detectors}
    The groundbreaking observation of GW150914 by the Advanced Laser Interferometer Gravitational-Wave Observatory (Advanced LIGO) in 2015 inaugurated GW astronomy and confirmed that coalescing stellar-mass black holes are prolific emitters of GWs \citep{abbottObservationGravitationalWaves2016}.  Terrestrial interferometers, however, are intrinsically limited below a few hertz by seismic, Newtonian-gravity and suspension-thermal noise, leaving the rich low-frequency band (0.1\,mHz–1\,Hz) inaccessible from the ground. To bridge this gap, a new generation of space-borne laser interferometers—such as LISA with its 2.5\,Gm triangular constellation \citep{colpiLISADefinitionStudy2024}, DECIGO featuring $\sim$1000\,km Fabry–Pérot arms, and the precursor concept to DECIGO, B-DECIGO (originally proposed as the “Pre-DECIGO” mission; \citealt{nakamuraPreDECIGOCanGet2016a})—will operate in drag-free flight and achieve picometre-level displacement sensitivity across the milli- to deci-hertz regime. Beyond enabling precision tests of strong-gravity astrophysics, these observatories are uniquely poised to detect the minute Doppler modulations caused by CBPs around BNS systems, thereby opening a novel gravitational-wave window onto the demographics and formation pathways of neutron-star exoplanets.

    The sensitivity curves are crucial for assessing the detectability of GW sources. The sensitivity of B-DECIGO, which depends on the GW frequency $f$, is described by \citep{isoyamaMultibandGravitationalWaveAstronomy2018}:
    \begin{equation}
    \begin{aligned}
        S_n^{\mathrm{B-DECIGO}}(f) = S_0\biggl[\,1.0 &+ 1.584 \times 10^{-2}\left(\frac{f}{1\,\mathrm{Hz}}\right)^{-4}\\
        &+ 1.584 \times 10^{-3}\left(\frac{f}{1\,\mathrm{Hz}}\right)^2\,\biggr] ,
    \end{aligned}
    \label{eq:snf b-decigo}
    \end{equation}
    where $S_0 = 4.040 \times 10^{-46}\,\mathrm{Hz}^{-1}$. The frequency range is $[f_{\text{low}},\; f_\text{high}] = [10^{-2},\;100] ~\mathrm{Hz}$.
    
    For DECIGO, the sensitivity is expressed as \citep{yagiDetectorConfigurationDECIGO2011}:
    \begin{equation}
    \begin{aligned}
        S_n^{\mathrm{DECIGO}}(f) =&\,7.05 \times 10^{-48}\left[1+\left(\frac{f}{f_p}\right)^2\right]\\
        &+4.8 \times 10^{-51}\left(\frac{f}{1\,\mathrm{Hz}}\right)^{-4}\frac{1}{1+\left(\frac{f}{f_p}\right)^2}\\
        &+5.33 \times 10^{-52}\left(\frac{f}{1\,\mathrm{Hz}}\right)^{-4}\,\mathrm{Hz}^{-1} ,
    \end{aligned}
    \label{eq:snf decigo}
    \end{equation}
    with $f_p=7.36\,\mathrm{Hz}$. The frequency range is $[f_{\text{low}},\; f_\text{high}] = [10^{-3},\;100] ~\mathrm{Hz}$.
    \begin{figure}
      \centering
      \includegraphics[width=\linewidth]{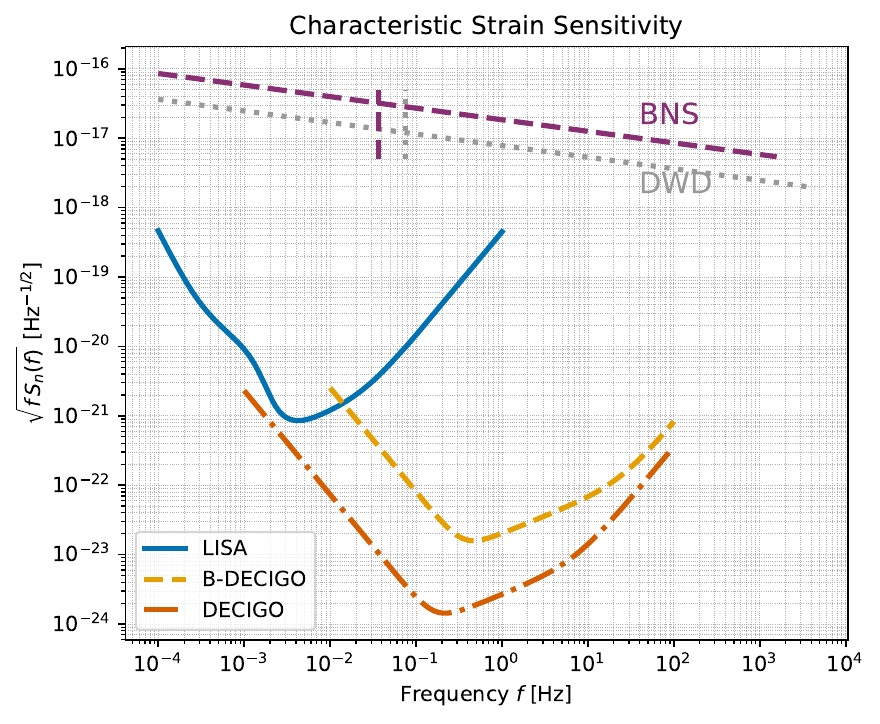}
      \caption{Detector sensitivity adopted in this work. We show the characteristic noise $h_n(f)\equiv\sqrt{f\,S_n(f)}$ for B-DECIGO and DECIGO. The two overlaid source tracks indicate the pre-merger signal strength of the reference compact binaries (BNS and DWD), and the vertical lines mark the GW frequencies reached $100\,\mathrm{years}$ before coalescence.}
      \label{fig:snf}
    \end{figure}

\subsection{CBP around BNS}    
    \begin{figure}
      \centering
      \includegraphics[width=\linewidth]{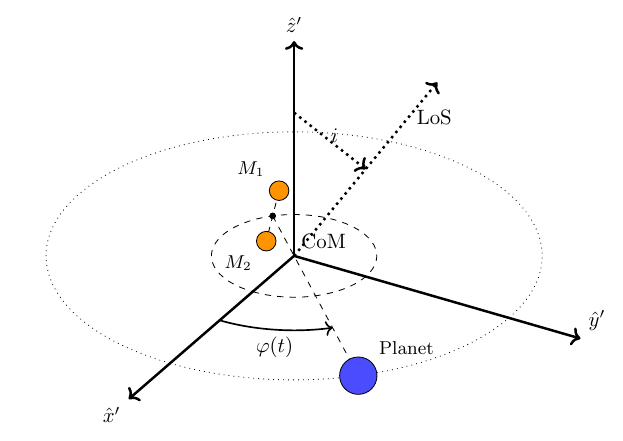}
      % \includestandalone[width=0.8\linewidth]{tikz/02_three_body_system}
      \caption{Illustration of a three-body system co
      mposed of a BNS and a CBP.}
      \label{fig:three-body system}
    \end{figure}

    We adopt the formalism presented in \cite{tamaniniGravitationalwaveDetectionExoplanets2019} to model the three-body system shown in Fig.~\ref{fig:three-body system}.  The source-frame coordinate system $\left(\hat{x}^{\prime},\; \hat{y}^{\prime},\; \hat{z}^{\prime}\right)$ is centered at the center of mass (CoM) of the three-body system. The two neutron stars, with masses $M_1$ and $M_2$, orbit their mutual CoM. The planet orbits the BNS CoM along a wider, circumbinary trajectory. The orbital phase of the planet is denoted by $\varphi(t)$. The observer's line of sight (LoS) lies at an inclination angle $i$ relative to the $\hat{z}^{\prime}$ axis.

    In the source reference frame, the position vector $\mathbf{r}(t)$ from the CoM of the BNS to the CBP is given by
    \begin{equation}
      \mathbf{r}(t) = \left(R\cos\varphi(t),\;R\sin\varphi(t),\;0\right) ,
    \end{equation}
    where the orbital radius $R$ of the planet is governed by Kepler's third law,
    \begin{equation}
      R^3 = (M_b + M_p)\left(\frac{P}{2\pi}\right)^2 ,
    \end{equation}
    and the orbital phase of the planet, $\varphi(t)$, evolves as
    \begin{equation}
      \varphi(t) = \frac{2\pi t}{P} + \varphi_0 ,
    \end{equation}
    with $M_b$, $M_p$, $P$, and $\varphi_0$ the total mass of the binary, the mass of the planet, the period and the initial phase of the planetary orbit, respectively.
    Since the planetary orbit is assumed to be circular, we can choose a source reference frame that is aligned with the $\hat{x}$-axis of the observation reference frame. Given the rotation matrix $\mathcal{R}(i)$ by the angle $i$ (the inclination angle between $\hat{z}'$ and the line of sight), we get the motion in the observation reference frame 
    \begin{equation}
    \begin{aligned}
        \mathcal{R}(i)\mathbf{r}
        &=\begin{pmatrix}
            1 & 0       & 0      \\
            0 & \cos i  & \sin i \\
            0 & -\sin i & \cos i 
        \end{pmatrix}
        \begin{pmatrix}
            R\cos\varphi(t) \\
            R\sin\varphi(t) \\
            0
        \end{pmatrix}\\
        &=(R\cos\varphi(t),\;R\cos i\sin\varphi(t),\;-R\sin i\sin\varphi(t)) .
    \end{aligned}
    \end{equation}
    The distance vector $\mathbf{r}_b(t)$ between the CoM of the three-body system and the CoM of the BNS is given by
    \begin{equation}
        \mathbf{r}_b(t)=\dfrac{M_p}{M_b+M_p}\mathcal{R}(i)\mathbf{r}(t) .
    \end{equation}
    The z-component of the motion is given by
    \begin{equation}
        z_b(t)=-\dfrac{M_p}{M_b+M_p}R\sin i \sin \varphi(t) .
    \end{equation}
    The velocity of the BNS CoM along the line of sight is then given by
    \begin{equation}
        v_{z,b}(t)=-\left(\frac{2 \pi}{P}\right)^{1/3} \frac{M_p}{\left(M_b+M_p\right)^{2/3}} \sin i  \cos \varphi(t).
    \end{equation}
    The parameter $K$ depends on the CBP mass $M_p$, its orbit inclination $i$, and the binary total mass $M_b$ given by 
    \begin{equation}
        K=\left(\frac{2 \pi G}{P}\right)^{1/3} \frac{M_p}{\left(M_b+M_p\right)^{2/3}} \sin i ,
    \end{equation}
    can simplify the velocity to
    \begin{equation}
        v_{z, b}(t)=-K \cos \varphi(t) .
    \end{equation}
    Since the BNS is in the early stable orbital phase, the frequency of the GW it emits can be considered to be stable at the frequency 
    \begin{equation}
        f_{\text{GW}}(t)=f_0+f_1 t+O\left(t^2\right),
    \end{equation}
    where $f_0$ is the initial observed frequency, $f_1$ given by
    \begin{equation}
        f_1=\frac{96}{5} \pi^{8 / 3} f_0^{11 / 3} \mathcal{M}^{5 / 3} ,
    \end{equation}
    is the time derivative of $f_\mathrm{GW}$ evaluated at the initial time and the higher order terms were neglected. $\mathcal{M}=\left(m_1 m_2\right)^{3 / 5} /\left(m_1+m_2\right)^{1 / 5}$ is the chirp mass. Due to the cosmological expansion, we measure the redshifted mass $m_{1,2}=(1+z) m_{1,2}^{\mathrm{S}}$ of the two compact objects, where $z$ is the redshift calculated from $D_L$, and $m_{1,2}^{\mathrm{S}}$ are the source-frame component masses with $m_1^{\mathrm{S}} \geq m_2^{\mathrm{S}}$ by default. Note that we include amplitude's
    Finally the GW frequency detected by the detector is given by
    \begin{equation}
        f_{\text{obs}}(t)=\left(1+\frac{v_{z, b}(t)}{c}\right) f_{\text{GW}}(t) .
    \end{equation}

\subsection{Waveform Construction}
    \begin{figure}
      \centering
      \includegraphics[width=\linewidth]{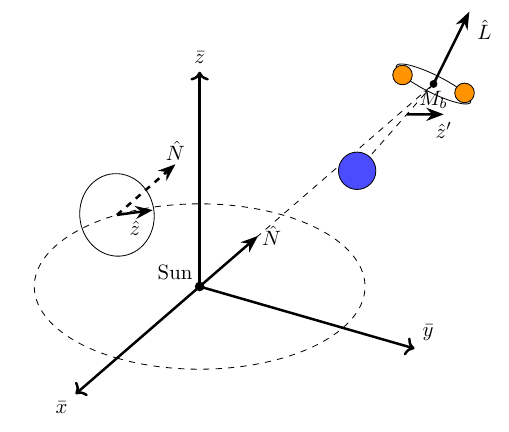}
      % \includestandalone[width=0.8\linewidth]{tikz/01_space_detector}
      \caption{Source–frame coordinates \((\hat{x}',\;\hat{y}',\;\hat{z}')\); 
               detector–frame coordinates \((\hat{x},\;\hat{y},\;\hat{z})\); 
               and ecliptic coordinates \((\bar{x},\;\bar{y},\;\bar{z})\).}
    \end{figure}

    We model the response of a space‐based interferometer with two independent data channels 
    \(\alpha = \mathrm{I},\mathrm{II}\) \citep{cutlerAngularResolutionLISA1998,
    tamaniniGravitationalwaveDetectionExoplanets2019}.  
    For a circular, Newtonian binary the measured strain is  
    \begin{equation}
      h_\alpha(t)
      =\frac{\sqrt{3}}{2}\,A_\alpha(t)\,\cos \chi_\alpha(t) ,
      \label{eq:strain}
    \end{equation}
    where the time–dependent amplitude is  
    \begin{equation}
      A_\alpha(t)=\bigl[A_{+}^{2}\,F^{+\,2}_{\alpha}(t)+A_{\times}^{2}\,
                    F^{\times\,2}_{\alpha}(t)\bigr]^{1/2} ,
      \label{eq:amp_total}
    \end{equation}
    and the total phase reads  
    \begin{equation}
      \chi_\alpha(t)=2\pi\!\int\! f_{\mathrm{obs}}\bigl(t'\bigr)\,dt'
                    +\Psi_0+\Phi_{p,\alpha}(t)+\Phi_D(t) ,
      \label{eq:phase_total}
    \end{equation}
    where $\Psi_0$ is a constant initial phase.
    
    The detector follows a heliocentric circular orbit of radius 
    \(R_{\mathrm{DEC}}=1\,\mathrm{AU}\) and period 
    \(T_{\mathrm{DEC}}=1\,\mathrm{year}\).  
    Its azimuthal position is  
    \begin{equation}
      \Phi(t)=\bar{\phi}_0+\frac{2\pi t}{T_{\mathrm{DEC}}}.
      \label{eq:detector_phi}
    \end{equation}
    With respect to the ecliptic frame \((\bar{x},\;\bar{y},\;\bar{z})\) the detector–frame basis vectors are  
    \begin{align}
      \hat{x}&=(-\sin\Phi,\;\cos\Phi,\;0) , \\
      \hat{y}&=\Bigl(-\frac12\cos\Phi,\;-\frac12\sin\Phi,\;-\frac{\sqrt{3}}{2}\Bigr) , \\
      \hat{z}&=\Bigl(-\frac{\sqrt{3}}{2}\cos\Phi,\;-\frac{\sqrt{3}}{2}\sin\Phi,\;
                    \frac12\Bigr) .
      \label{eq:basis_vectors}
    \end{align}

    The line–of–sight unit vector \(\hat{N}\) and the binary’s orbital–angular‐momentum unit
    vector \(\hat{L}\) are written in the ecliptic frame as  
    \begin{align}
      \hat{N}&=\bigl(\sin\bar{\theta}_S\cos\bar{\phi}_S,\;
                     \sin\bar{\theta}_S\sin\bar{\phi}_S,\;
                     \cos\bar{\theta}_S\bigr) , \\
      \hat{L}&=\bigl(\sin\bar{\theta}_L\cos\bar{\phi}_L,\;
                     \sin\bar{\theta}_L\sin\bar{\phi}_L,\;
                     \cos\bar{\theta}_L\bigr) .
      \label{eq:N_L_vectors}
    \end{align}
    Their instantaneous sky position and polarization in the detector frame are  
    \begin{align}
      \cos\theta &= \hat{N}\!\cdot\!\hat{z} , \label{eq:theta}\\
      \phi &= \arctan\!\left(\frac{\hat{N}\!\cdot\!\hat{y}}
                                   {\hat{N}\!\cdot\!\hat{x}}\right)
              +\frac{2\pi t}{T_{\mathrm{DEC}}}+\alpha_0 , \label{eq:phi}\\
      \tan\psi &=\frac{\hat{L}\!\cdot\!\hat{z}
                     -(\hat{L}\!\cdot\!\hat{N})(\hat{z}\!\cdot\!\hat{N})}
                    {\hat{N}\!\cdot\!\bigl(\hat{L}\times\hat{z}\bigr)} ,
      \label{eq:psi}
    \end{align}
    where the angle $\alpha_0$ is the initial orientation of the detector arms.
    
    With \((\theta,\;\phi,\;\psi)\) defined above, the detector response functions are  
    \begin{align}
      F_{\mathrm{I}}^{+} &= 
        \frac12\bigl(1+\cos^{2}\theta\bigr)\cos2\phi\cos2\psi
        -\cos\theta\sin2\phi\sin2\psi , \\
      F_{\mathrm{I}}^{\times} &= 
        \frac12\bigl(1+\cos^{2}\theta\bigr)\cos2\phi\sin2\psi
        +\cos\theta\sin2\phi\cos2\psi , \\
      F_{\mathrm{II}}^{+,\times} &=
        F_{\mathrm{I}}^{+,\times}\!\left(\theta,\;\phi-\pi/4,\;\psi\right) .
      \label{eq:antenna}
    \end{align}

    The plus and cross amplitudes are  
    \begin{align}
      A_{+} &= 2\mathcal{A}\Bigl[1+\bigl(\hat{L}\!\cdot\!\hat{N}\bigr)^{2}\Bigr] , \\
      A_{\times} &= -4\mathcal{A}\bigl(\hat{L}\!\cdot\!\hat{N}\bigr) ,
    \end{align}
    where the overall amplitude scale is  
    \begin{equation}
      \mathcal{A}=\frac{\mathcal{M}^{5/3}\bigl(\pi f_0\bigr)^{2/3}}{D_L} .
      \label{eq:calA}
    \end{equation}
    
    The polarization phase is  
    \begin{equation}
      \Phi_{p,\alpha}(t)=\arctan\!\left(
        -\,\frac{A_{\times}F_{\alpha}^{\times}(t)}{A_{+}F_{\alpha}^{+}(t)}
      \right) ,
    \end{equation}
    and the detector Doppler phase is  
    \begin{equation}
      \Phi_D(t)=2\pi f_{\mathrm{obs}}(t)\,R_{\mathrm{DEC}}\,
                \sin\bar{\theta}_S\,
                \cos\!\bigl[\Phi(t)-\bar{\phi}_S\bigr].
    \end{equation}

    Each GW signal from a BNS in a CBP system is thus characterized by  
    \[
      \boldsymbol{\lambda}=
      \bigl\{\ln\mathcal{A},\;\Psi_0,\;f_0,\;f_1,\;
             \bar{\theta}_S,\;\bar{\phi}_S,\;
             \bar{\theta}_L,\;\bar{\phi}_L,\;
             K,\;P,\;\varphi_0\bigr\} .
    \]
    Here \((\bar{\theta}_S,\;\bar{\phi}_S)\) locate the source on the sky,  
    \((\bar{\theta}_L,\;\bar{\phi}_L)\) fix the orbital orientation,  
    and the remaining parameters describe the intrinsic CBP–BNS dynamics 
    and the emitted GW.

\subsection{Parameter estimation}
\label{subsec:parameter_estimation}
    Following \cite{tamaniniGravitationalwaveDetectionExoplanets2019}, we compute the signal-to-noise ratio (SNR) of the GW signal as
    \begin{equation}
    \mathrm{SNR}^2=\frac{2}{S_n\!(f_0)}\sum_{\alpha=\mathrm{I,II}}\int_0^{T_{\mathrm{obs}}}\!h_\alpha^2(t)\,\mathrm{d}t\,,
    \end{equation}

    the observational time is set to $T_{\mathrm{obs}} = 3\,\mathrm{year}$ for B-DECIGO and $T_{\mathrm{obs}} = 4\,\mathrm{years}$ for DECIGO. For DECIGO, which consists of four identical interferometers operating simultaneously, an additional factor of $4$ is included in the above expression to account for the enhanced combined sensitivity; this factor is also applied when computing the Fisher matrix in subsequent analyses. The one-sided noise power spectral densities $S_n(f)$ employed in this work are given by Eqs.~\eqref{eq:snf b-decigo} and~\eqref{eq:snf decigo}.

    The uncertainties and correlations among the estimated parameters can be derived using the covariance matrix, which is the inverse of the Fisher information matrix:
    \begin{equation}
        \Sigma_{ij}=\left\langle\Delta\lambda_i\Delta\lambda_j\right\rangle=(\Gamma^{-1})_{ij} ,
    \end{equation}
    with the Fisher information matrix $\Gamma_{ij}$ given by
    \begin{equation}
        \Gamma_{ij}=\frac{2}{S_n(f_0)} \sum_{\alpha=\mathrm{I},\mathrm{II}} \int_0^{T_{\mathrm{obs}}}\frac{\partial h_\alpha(t)}{\partial \lambda_i}\frac{\partial h_\alpha(t)}{\partial \lambda_j}\,\mathrm{d}t .
    \end{equation}
    For GW signals with sufficiently high SNR, the statistical uncertainty on each parameter $\lambda_i$ can be approximated as $\sqrt{\Sigma_{ii}}$.

    Adopting the approximation used in \cite{cutlerAngularResolutionLISA1998}, we rewrite the Fisher matrix as
    \begin{equation}
    \begin{aligned}
        \Gamma_{ij} =&\; S_n^{-1}(f_0) \sum_{\alpha=\mathrm{I},\mathrm{II}} \int_{0}^{T_{\mathrm{obs}}}\left[\frac{\partial A_\alpha(t)}{\partial \lambda_i}\frac{\partial A_\alpha(t)}{\partial \lambda_j}\right. \\
        &\quad \left.+ A_\alpha^2(t)\frac{\partial \chi_\alpha(t)}{\partial \lambda_i}\frac{\partial \chi_\alpha(t)}{\partial \lambda_j}\right]\,\mathrm{d}t .
    \end{aligned}
    \end{equation}

    The partial derivatives with respect to the angular parameters $(\bar{\theta}_S, \bar{\phi}_S, \bar{\theta}_L, \bar{\phi}_L)$ are evaluated numerically, whereas the remaining analytical derivatives are
    \begin{equation}
    \begin{aligned}
        & \frac{\partial A_\alpha(t)}{\partial \ln \mathcal{A}}=A_\alpha(t), \quad \frac{\partial A_\alpha(t)}{\partial \Psi_0}=0, \quad \frac{\partial A_\alpha(t)}{\partial f_0}=0,\\
        & \frac{\partial A_\alpha(t)}{\partial f_1}=0, \quad \frac{\partial A_\alpha(t)}{\partial K}=0, \quad \frac{\partial A_\alpha(t)}{\partial P}=0, \quad \frac{\partial A_\alpha(t)}{\partial \varphi_0}=0,\\
        & \frac{\partial \chi_\alpha(t)}{\partial \ln \mathcal{A}}=0, \quad \frac{\partial \chi_\alpha(t)}{\partial \Psi_0}=1, \quad \frac{\partial \chi_\alpha(t)}{\partial f_0}=2 \pi t - K P \sin \varphi(t),\\
        & \frac{\partial \chi_\alpha(t)}{\partial f_1}=\pi t^2 - K P t \sin \varphi(t)-\frac{K P^2}{2 \pi} \cos \varphi(t),\\
        & \frac{\partial \chi_\alpha(t)}{\partial K}=-f_0 P \sin \varphi(t),\quad \frac{\partial \chi_\alpha(t)}{\partial \varphi_0}=-f_0 K P \cos \varphi(t),\\
        & \frac{\partial \chi_\alpha(t)}{\partial P}=-f_0 K \sin \varphi(t)+\frac{2 \pi f_0 K t}{P} \cos \varphi(t)\,.
    \end{aligned}
    \end{equation}

    Following the criterion set in \cite{tamaniniGravitationalwaveDetectionExoplanets2019}, we consider a planet detection significant when the fractional uncertainties on both parameters $K$ and $P$ are below $30\%$.
    
\section{Result}\label{sec:result}
    In our analysis, we set two reference GW sources—a BNS and a DWD—to systematically evaluate the detection capability of B-DECIGO and DECIGO for CBPs around compact binaries. By varying individual source parameters from these baseline settings, we also investigate how each parameter influences the detector's sensitivity and capability to detect the CBP-BNS systems.
    
    Following similar parameter setups adopted in previous LISA studies \cite{kangProspectsDetectingExoplanets2021}, we fix the sky location of our sources at ecliptic coordinates $\bar{\theta}_S=1.27$, $\bar{\phi}_S=5$, and a luminosity distance $D_L=10\,\mathrm{kpc}$. The orbital orientation of the source binary system is fixed at $\bar{\theta}_L=0.5$, $\bar{\phi}_L=3.981$, with the constant initial GW phase set as $\Psi_0=0$. For the CBP, we adopt an initial orbital phase $\varphi_0=\pi/2$ and an inclination angle $i=\pi/3$.
    
    In our default scenario for the BNS system, we consider equal component masses of $m_1 = m_2 = 1.4\,M_\odot$, similar to the GW170817 event observed by the LIGO–Virgo Collaboration \citep{collaborationGW170817ObservationGravitational2017}, with an initial GW frequency $f_0 = 0.05\,\mathrm{Hz}$. This configuration yields a chirp mass of $\mathcal{M} \simeq 1.22\,M_\odot$ and an inspiral timescale of $\sim 5\times10^2$\,years, indicating that the binary evolves adiabatically during the 3–4\,years observation and remains in the early inspiral stage. For the DWD system, we set the component masses to $m_1 = m_2 = 0.5\,M_\odot$, with an initial GW frequency $f_0 = 0.005\,\mathrm{Hz}$, corresponding to an inspiral timescale of $\sim 2\times10^5$\,years for a circular binary with chirp mass $\mathcal{M} = 0.435\,M_\odot$. Hence the DWD is well detached throughout the mission duration, and no Roche-lobe overflow or merger is expected. These timescales confirm that both fiducial binaries can be treated as quasi-monochromatic sources on the B-DECIGO and DECIGO observing baselines.

    These reference parameters provide a robust baseline for exploring the feasibility and sensitivity of B-DECIGO and DECIGO detectors in identifying CBP signatures around compact binary systems, while systematically assessing how deviations from these parameters influence detection performance.

\subsection{Planetary Parameter Estimation}
    \begin{figure}
      \centering
      \includegraphics[width=\linewidth]{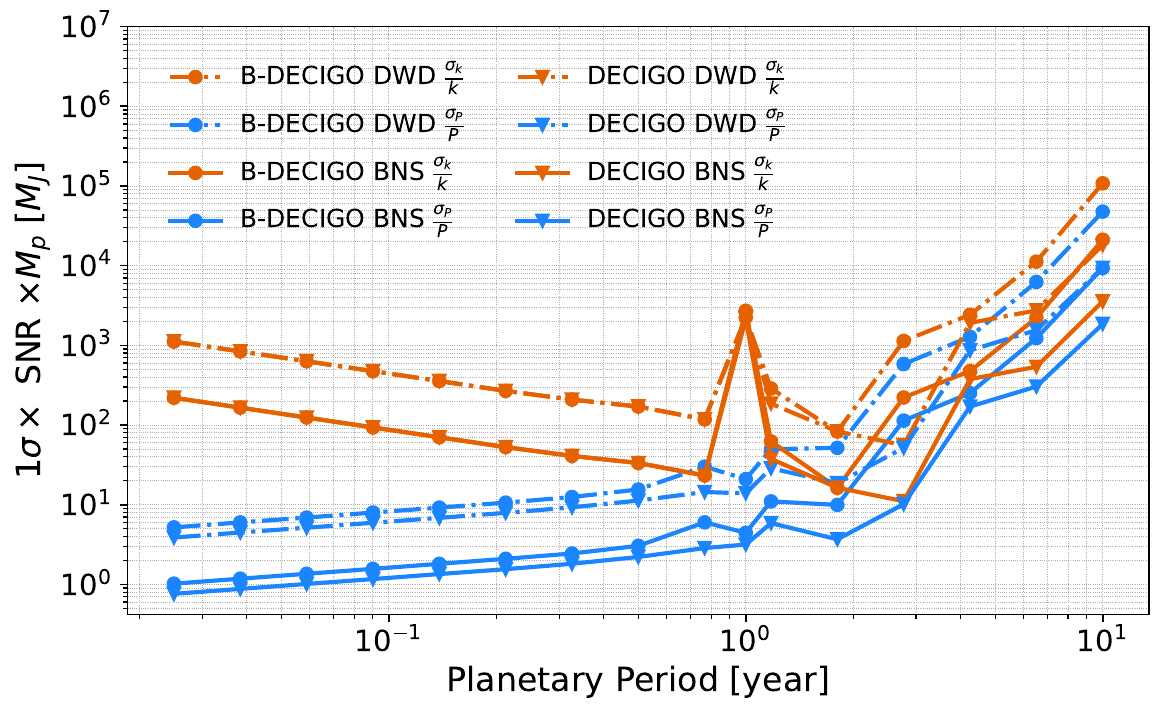}
      \caption{Normalized 1$\sigma$ errors on the circumbinary-planet parameters as a function of the planetary period $P$ for BNS and DWD sources observed by B-DECIGO and DECIGO. The curves show the error on $K$ and on $P$, each rescaled by the GW SNR and by $M_p$ (in $M_J$), so that lower values indicate tighter constraints. Errors remain small when multiple orbital cycles are sampled, develop a minimum at intermediate $P$ (where the Doppler imprint strengthens), and then rise sharply once $P$ approaches or exceeds the mission duration; a feature near $P\!\sim\!1\,\mathrm{year}$ reflects degeneracy with the detector’s annual motion. Default source and observing assumptions are used throughout.}
      \label{fig:decigo_comparison}
    \end{figure}
    The normalized parameter--estimation errors—defined as the product of the $1\sigma$ uncertainty, the signal--to--noise ratio, and the planetary mass $M_p$ expressed in Jupiter masses—for CBPs orbiting BNS and DWD systems measured with both B-DECIGO and the full DECIGO mission are displayed in Fig.~\ref{fig:decigo_comparison}.  For short planetary periods the errors remain low for every configuration, because several complete orbits are sampled within the observing time, allowing the gravitational--wave Doppler modulation produced by the planet to accumulate cleanly.
    
    As the period grows, each curve develops an inflection point. To the left of this turning point a modest increase in period can reduce the error, since a wider planetary orbit generates a larger barycentric signal. Beyond the turning point the errors rise steeply: once the orbital period approaches or exceeds the total observing time, only a fraction of the orbit is tracked and the planetary frequency modulation becomes degenerate with the source’s intrinsic frequency evolution and with instrumental noise.
    
    B-DECIGO diverges from DECIGO at long periods. The dominant driver is the shorter nominal mission lifetime of B-DECIGO (3~years versus 4~years for DECIGO): a briefer baseline leaves a longer--period planet less well sampled, which inflates the post--normalization uncertainties. The source class modulates the effect: BNS signals, with their higher GW frequencies and intrinsic chirps, preserve the planetary Doppler imprint more effectively than the relatively quiescent DWDs, so the error blow‐up sets in later for BNS systems.
    
    Overall, B-DECIGO retains a comparative advantage for high‐frequency BNS systems at periods below $\simeq 1$~year, while DECIGO’s longer observing window allows it to maintain tighter constraints at the multi‐year end of the period range.  These combined results corroborate earlier LISA‐based studies\citep{tamaniniGravitationalwaveDetectionExoplanets2019}: optimal sensitivity and precision are obtained for planetary periods of a few years or less, whereas longer periods suffer a rapid loss of fidelity—here shown to depend sensitively on the total mission duration as well as on the intrinsic properties of the GW source.

\subsection{Selection Function}
    \begin{figure}
      \centering
      \includegraphics[width=\linewidth]{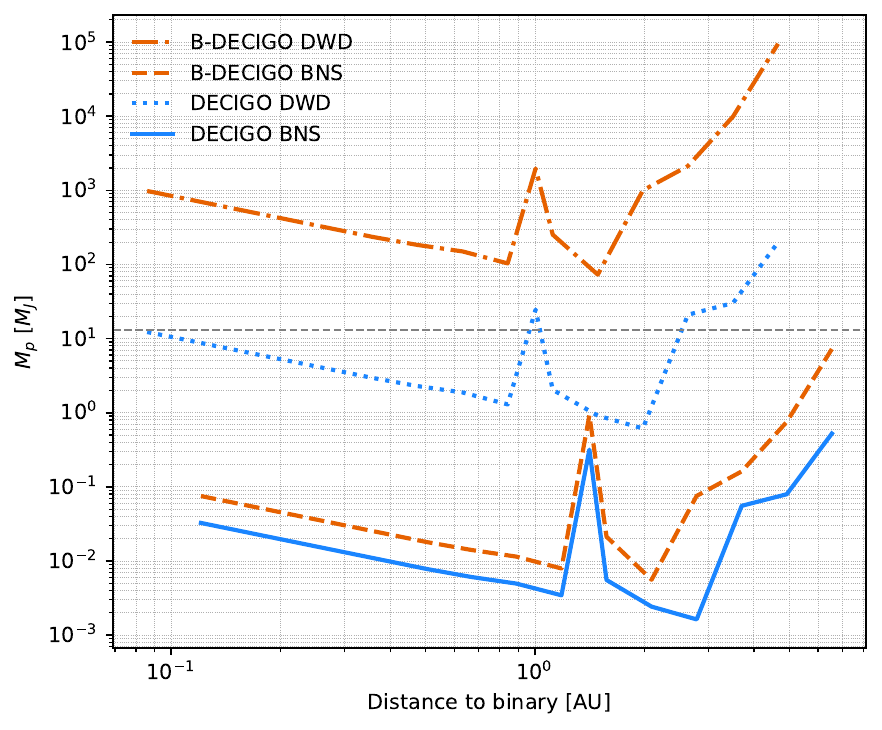}
      \caption{Selection function in the plane of planetary mass versus orbital separation for CBPs around compact binaries. The solid curves give, for each detector and source class (BNS/DWD), the minimum detectable mass $M_p^{\rm min}$ at a given orbital distance $a$, adopting the detection criterion $\sigma_K/K<0.3$ and $\sigma_P/P<0.3$. The horizontal dashed line marks the deuterium-burning limit at $13\,M_J$. DECIGO outperforms B-DECIGO across the board, and BNS hosts yield substantially lower $M_p^{\rm min}$ than DWDs due to their higher carrier GW frequencies.}
      \label{fig:decigo_curve}
    \end{figure}
    
    Fig.~\ref{fig:decigo_curve} presents the simulated results based on previously established default parameters, utilizing a detection criterion requiring a relative uncertainty better than 30\%. This figure illustrates the detection capabilities of space-based gravitational-wave detectors for CBPs orbiting compact binary systems. Specifically, the horizontal axis represents the orbital distance of the planet from the binary system’s center of mass (in astronomical units, AU), while the vertical axis denotes the minimum detectable planetary mass (in Jupiter masses, $M_J$), given the stated uncertainty criterion. The solid curves show the comparative sensitivities of the B-DECIGO and LISA detectors BNS and DWD systems.
    
    The results distinctly highlight the superior performance of B-DECIGO over LISA across nearly the entire orbital range, especially in the detection of planets orbiting BNS systems. This improved sensitivity originates from B-DECIGO's operational frequency band, which closely matches the higher-frequency gravitational-wave signals typically emitted by BNS sources, making it highly effective in identifying the subtle Doppler frequency modulation caused by planetary perturbations. Remarkably, B-DECIGO demonstrates the capability to detect planets down to approximately Earth-mass levels around BNS systems, significantly exceeding LISA’s sensitivity and enabling potential discoveries of extremely low-mass exoplanets within our Galaxy. Because DECIGO is the full, mature implementation of the B-DECIGO concept, its curves lie everywhere below the red B-DECIGO curves for both BNS and DWD sources, signaling uniformly better sensitivity.

    The horizontal dashed line marks the 13 Jupiter-mass threshold ($13\,M_J$), traditionally adopted as the dividing criterion between planets and brown dwarfs, corresponding to the critical mass required for deuterium burning \citep{tamaniniGravitationalwaveDetectionExoplanets2019}. Objects below this mass are conventionally classified as genuine planets, whereas those exceeding it are considered brown dwarfs. According to the presented results, DECIGO pushes the minimum detectable mass for CBPs around DWD systems well below the $13 M_J$ planet–brown‐dwarf boundary across a broad span of orbital separations, whereas LISA dips beneath this threshold only in a narrow inner zone and otherwise stays at or above the $13\,M_J$ limit. Conversely, the detection capabilities for BNS systems extend significantly further outwards, with B-DECIGO/DECIGO capable of detecting planets far below the planetary/brown dwarf threshold across a wide range of orbital distances, clearly underscoring its superior performance.
    
    In terms of general trends, the minimum detectable planet mass does not monotonically increase with orbital distance. Instead, it initially decreases at smaller orbital distances, reflecting reduced detectability due to weaker Doppler modulation effects when the planetary orbit is too close to the binary system. As the orbital distance further increases, there is a pronounced rise in the detection mass limit, 
    arising from a degeneracy that occurs when the planetary period approaches the detector’s one-year heliocentric motion ($\sim1\,\mathrm{year}$), 
    which substantially degrades the signal-to-noise ratio. At even larger orbital separations, the minimum detectable mass limit gradually increases again, as extremely long orbital periods exceed the observational time span, resulting in incomplete orbital coverage and consequently insufficient signal strength for reliable parameter estimation. These features reveal the nuanced interplay between orbital configurations and detection sensitivity, emphasizing the exceptional potential of B-DECIGO/DECIGO for exploring planetary systems around compact binaries, particularly BNS systems, thus significantly enhancing our understanding of planetary formation and evolution processes in extreme astrophysical environments.

\subsection{Effects of \texorpdfstring{$D_L$}{DL}, Masses and, \texorpdfstring{$f_0$}{f0}}
    \begin{figure}
      \centering
      \includegraphics[width=\linewidth]{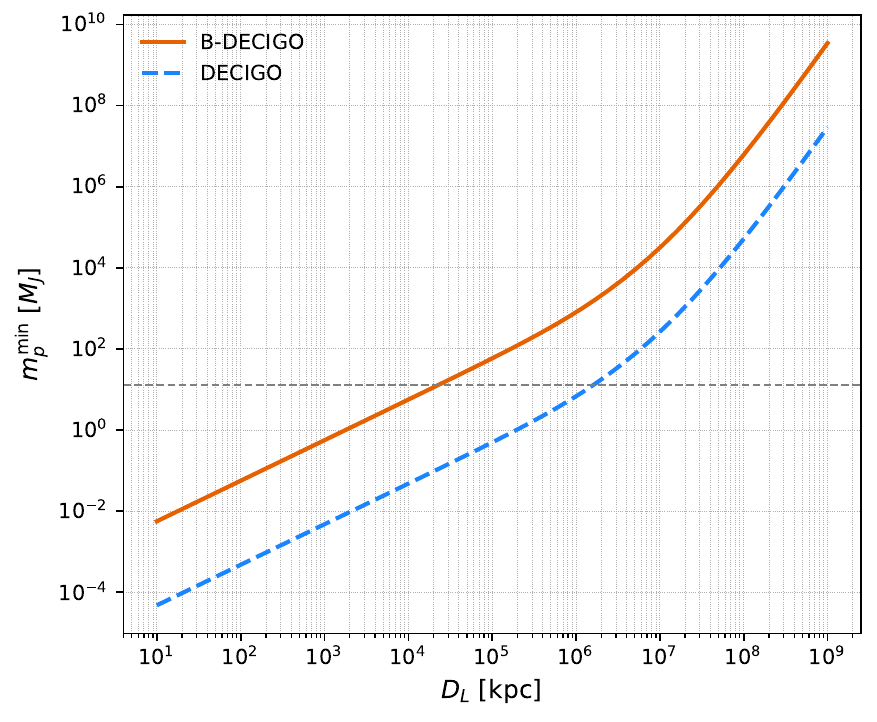}
      \caption{Minimum detectable planetary mass $M_p^{\rm min}$ versus luminosity distance $D_L$ for equal-mass BNS reference systems, comparing B-DECIGO (3\,years) and DECIGO (4\,year). As $D_L$ increases the SNR drops and $M_p^{\rm min}$ rises; DECIGO’s lower noise floor and longer integration push the sensitivity from super-Earth/Neptune scales in the Galaxy to sub-Jovian levels well beyond, under the same 30\% detection criterion on $(K,P)$.}
      \label{fig:result_DL}
    \end{figure}
    
    Fig.~\ref{fig:result_DL} shows the detection capability of B-DECIGO/DECIGO for CBPs orbiting compact binaries at various luminosity distances $D_L$. As $D_L$ increases, the minimum detectable planet mass $m_p^{\min }$ rises notably. This effect results from the significant attenuation of GW signal strength over greater distances, drastically lowering the SNR and necessitating increasingly massive planets to produce detectable Doppler modulation signals. Despite the improved sensitivity of detectors such as B-DECIGO/DECIGO compared to LISA, the capability to detect planets inevitably decreases with distance. Within the Milky Way halo and its immediate neighbourhood DECIGO drives the limit down to $\sim10^{-4}\,M_J$, corresponding to Earth‐mass planets, whereas B-DECIGO remains at the few‐Earth‐mass level ($\sim10^{-2}\,M_J$). This dramatic improvement is a direct consequence of DECIGO’s lower noise floor and longer mission lifetime.

    \begin{figure}
      \centering
      \includegraphics[width=\linewidth]{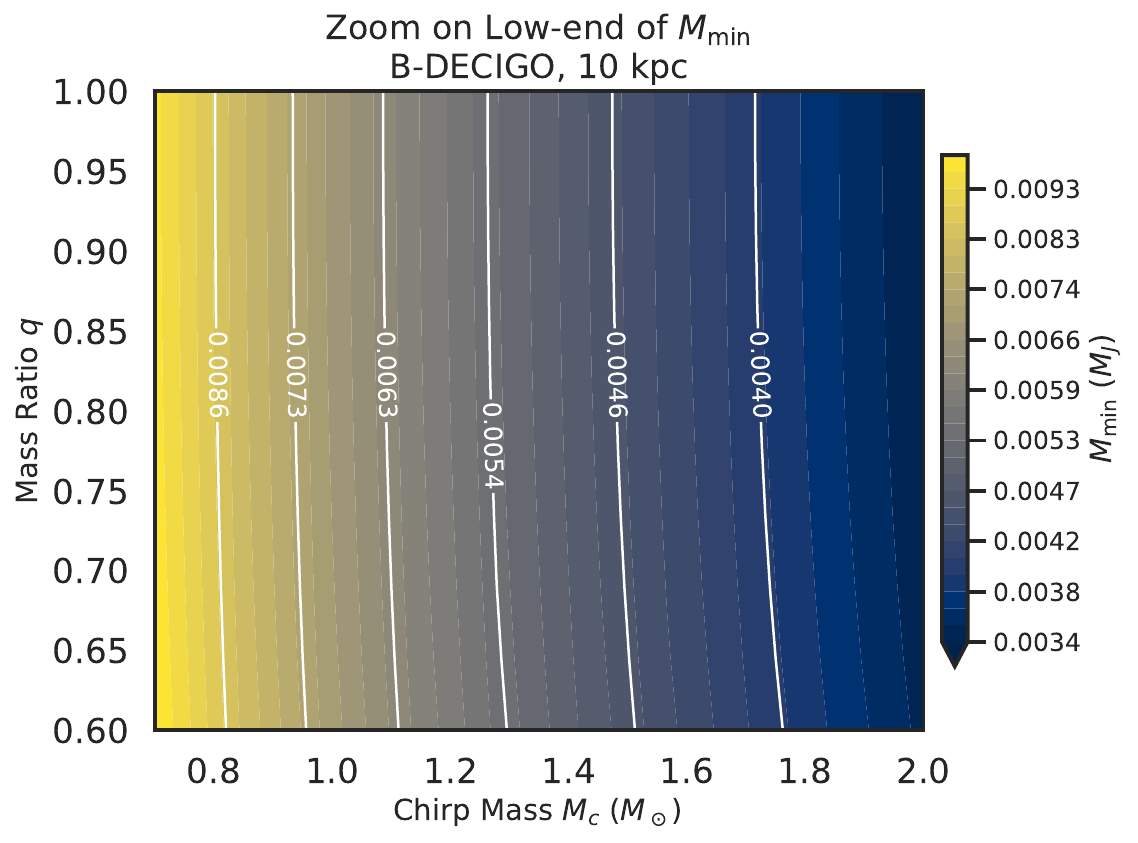}
      \caption{B-DECIGO: dependence of the minimum detectable planetary mass $M_p^{\rm min}$ on the binary chirp mass $M_c$ and component-mass ratio $q$ at $D_L=10\,\mathrm{kpc}$ (Galactic case). Contours show that higher $M_c$ and more equal masses ($q\!\to\!1$) significantly improve sensitivity, driving $M_p^{\rm min}$ down to the super-Earth regime for favorable BNS-like systems.}
      \label{fig:result_mass}
    \end{figure}

    \begin{figure}
      \centering
      \includegraphics[width=\linewidth]{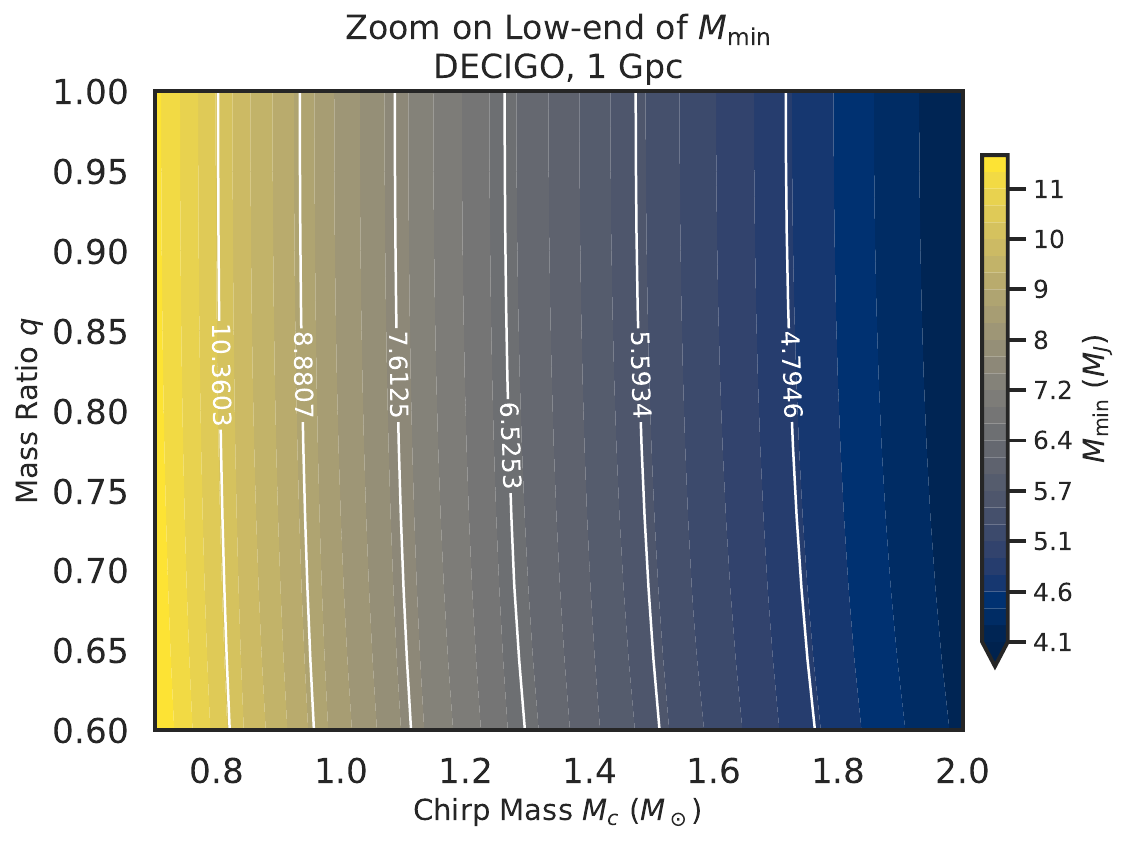}
      \caption{DECIGO: same as Fig.~7 but for an extragalactic case at $D_L=1\,\mathrm{Gpc}$. Even at cosmological distances, high-$M_c$, near-equal-mass binaries allow detection of giant planets with $M_p^{\rm min}$ of a few Jupiter masses under the adopted criterion and mission duration.}
      \label{fig:result_mass_decigo}
    \end{figure}

    Figures~\ref{fig:result_mass} and~\ref{fig:result_mass_decigo} show the dependence of the minimum detectable planetary mass $m_{\min}$ on the binary’s chirp mass $M_c$ and component mass ratio $q$ for B-DECIGO and DECIGO under two observational scenarios. In the B-DECIGO case (Fig.~\ref{fig:result_mass}), the source is placed at $D_L = 10\,\mathrm{kpc}$, representing binaries within the Milky Way, while in the DECIGO case (Fig.~\ref{fig:result_mass_decigo}), the source is located at $D_L = 1\,\mathrm{Gpc}$, corresponding to binaries in distant galaxies. In both situations, higher chirp masses and nearly equal component masses ($q \approx 1$) enable the detection of significantly smaller planets. Larger chirp masses produce stronger GW signals, increasing the SNR, while nearly equal-mass binaries maximize the gravitational-wave amplitude. This combination enhances the planetary Doppler signature, making low-mass planets easier to distinguish from the intrinsic binary waveform. For the Galactic scenario, B-DECIGO achieves extreme sensitivity, with $m_{\min}$ reaching ${\sim}4\times 10^{-3}\,M_J$ for $M_c \gtrsim 1.8\,M_\odot$ and $q \approx 1$, corresponding to super-Earth masses. In the extragalactic DECIGO case at $1\,\mathrm{Gpc}$, the sensitivity remains at the giant-planet level, with $m_{\min}$ as low as ${\sim}4\,M_J$ for similar binary parameters—remarkable given the three orders of magnitude greater distance. The contour distributions in both figures indicate that increasing $M_c$ yields more pronounced sensitivity gains than equivalent changes in $q$, although both parameters act synergistically to reduce $m_{\min}$. These results highlight that B-DECIGO is best suited for detecting sub-Jovian planets within the Galaxy, while DECIGO extends the search to giant planets orbiting compact binaries in galaxies billions of light-years away, well beyond the Local Supercluster, enabling a cosmological-scale survey of CBPs.

    \begin{figure}
      \centering
      \includegraphics[width=\linewidth]{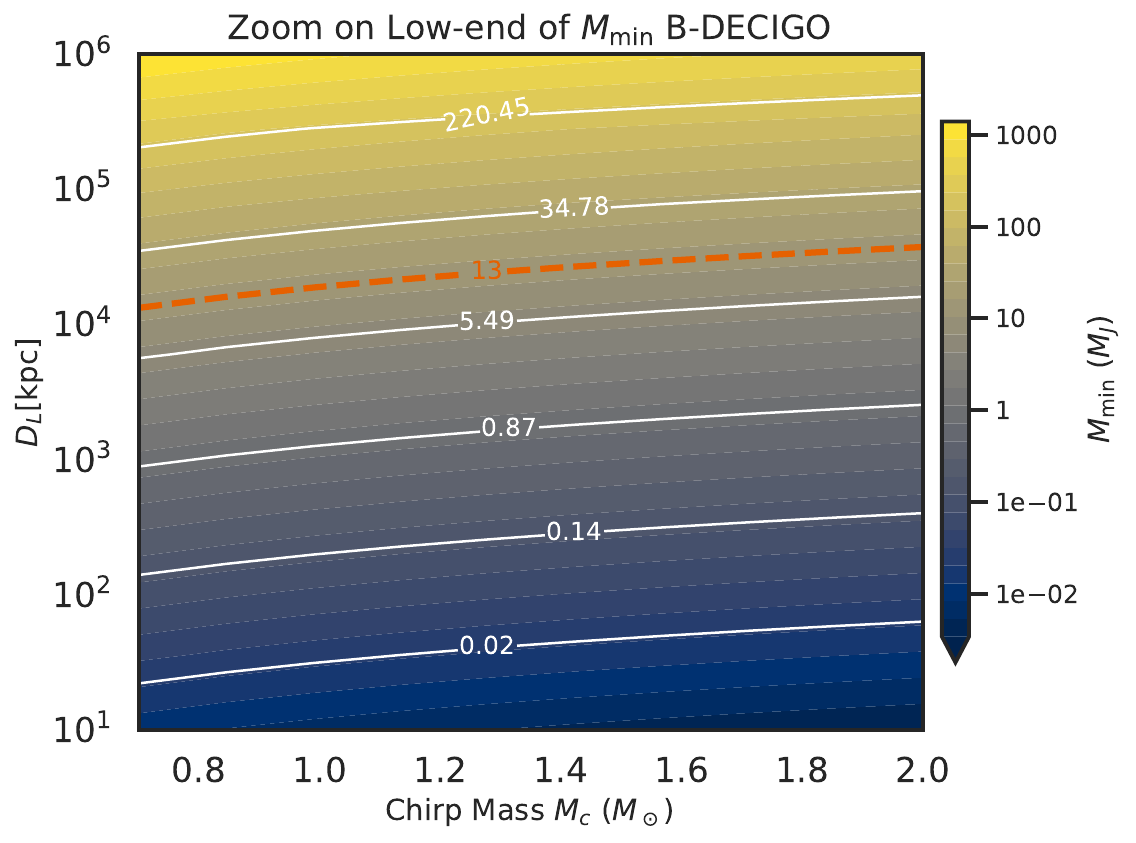}
      \caption{B-DECIGO: joint dependence of $M_p^{\rm min}$ on the binary chirp mass $M_c$ and luminosity distance $D_L$. At fixed $D_L$, increasing $M_c$ (hence the intrinsic GW amplitude) lowers $M_p^{\rm min}$; at fixed $M_c$, increasing $D_L$ degrades sensitivity. The combination highlights massive, nearby binaries as prime targets for sub-Jovian and even sub-Neptune planets.}
      \label{fig:result_mass_DL}
    \end{figure}

    \begin{figure}
      \centering
      \includegraphics[width=\linewidth]{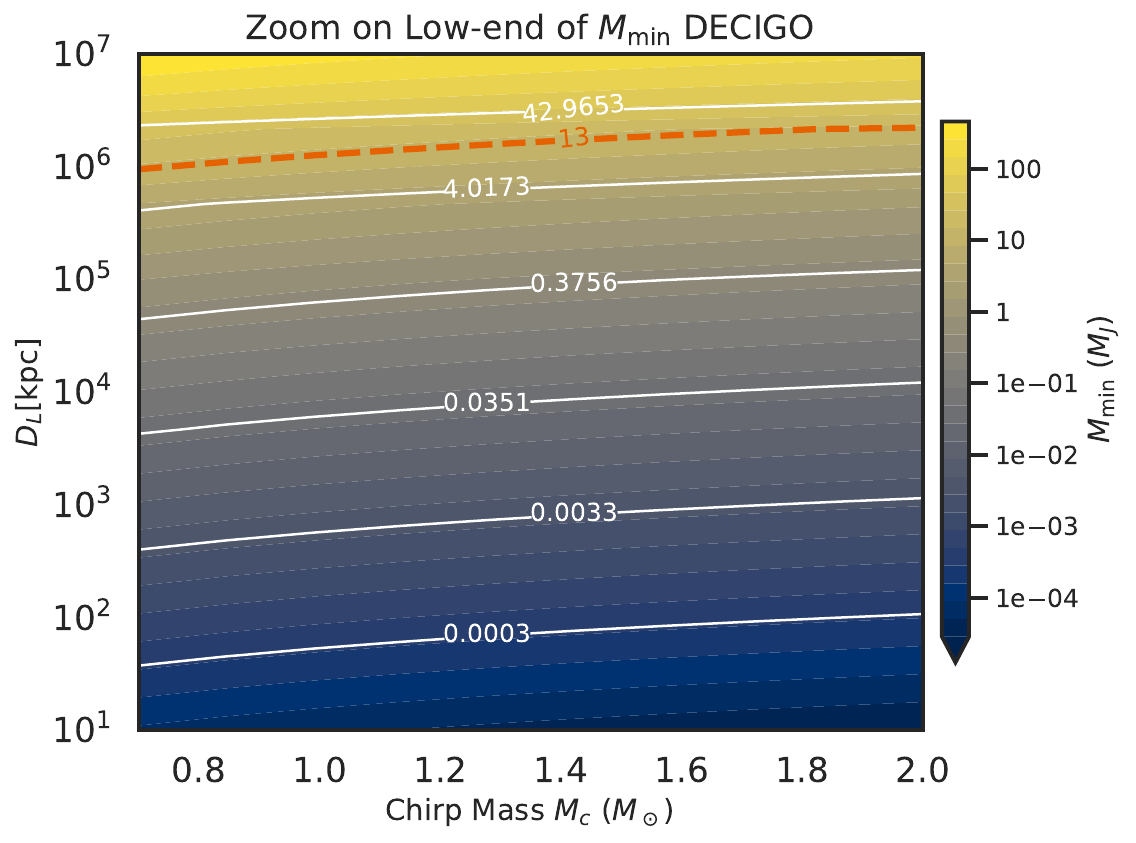}
      \caption{DECIGO: same as Fig.~9. Relative to B-DECIGO, DECIGO achieves uniformly lower $M_p^{\rm min}$ across the $(M_c,D_L)$ plane—extending Earth-mass sensitivity to nearby sources and maintaining sub-Jovian reach to distances orders of magnitude larger.}
      \label{fig:result_mass_DL_decigo}
    \end{figure}
    
    Figures~\ref{fig:result_mass_DL} and~\ref{fig:result_mass_DL_decigo} illustrate how the minimum detectable planetary mass $m_{\min}$ depends jointly on the chirp mass $M_c$ of the binary and the luminosity distance $D_L$ for B-DECIGO and DECIGO. In both cases, $m_{\min}$ decreases systematically with increasing $M_c$ at any fixed $D_L$, reflecting the fact that more massive binaries emit stronger gravitational waves and thus achieve higher intrinsic signal-to-noise ratios, which in turn allows the detection of lower-mass planets. This trend is particularly pronounced for nearly equal-mass, high-$M_c$ systems such as binary neutron stars, where the gravitational-wave amplitude is maximised and the planetary Doppler signature is more readily distinguished from the intrinsic binary signal. While the overall rise of $m_{\min}$ with distance has been discussed earlier, the present figures highlight that the rate of degradation with $D_L$ is strongly modulated by $M_c$: high-$M_c$ systems retain sensitivity to low-mass planets over much greater distances, whereas low-$M_c$ binaries rapidly lose the ability to detect small companions as $D_L$ increases. Comparing the two detectors, DECIGO achieves uniformly lower $m_{\min}$ across the entire $(M_c, D_L)$ parameter space, extending Earth-mass sensitivity to nearby sources and maintaining sub-Jovian detection capability to distances orders of magnitude beyond those accessible to B-DECIGO.

    \begin{figure}
      \centering
      \includegraphics[width=\linewidth]{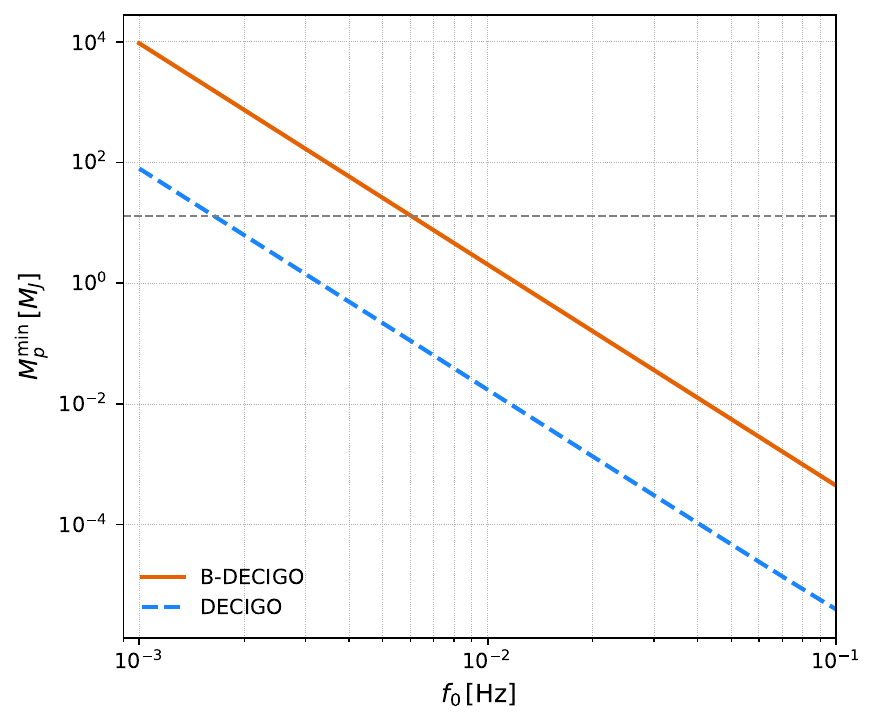}
      \caption{Minimum detectable planetary mass $M_p^{\rm min}$ versus the source’s initial GW frequency $f_0$ for the reference BNS systems. Sensitivity improves steeply with $f_0$, since the Doppler sidebands scale with the carrier frequency and the SNR. DECIGO can reach Earth-mass scales for high-$f_0$ BNSs in the Milky Way, while B-DECIGO remains sensitive to super-Earth/Neptune analogues under the same assumptions.}
      \label{fig:result_f0}
    \end{figure}
    Fig.~\ref{fig:result_f0} demonstrates the substantial influence of initial orbital frequency $f_0$ on B-DECIGO/DECIGO's sensitivity to CBPs. For both detectors the mass threshold falls steeply with increasing $f_0$: binaries that already emit gravitational waves at higher frequencies produce stronger carrier signals and larger planet‐induced Doppler sidebands, thereby boosting the signal–to–noise ratio (SNR) and enabling the detection of progressively lower‐mass companions. Additionally, the planetary Doppler-induced frequency shifts are more prominent at higher carrier frequencies, facilitating clearer extraction from observational data. Moreover, higher-frequency systems effectively mitigate interference arising from resonance effects associated with the detector's own orbital period modulation, significantly boosting sensitivity to smaller planets.
    
    High–frequency BNS systems present the most favorable targets: DECIGO could detect Earth‐mass planets—and potentially large moons—around such binaries throughout the Milky Way, while  B-DECIGO remains sensitive to super‐Earth and Neptune analogues.  At lower $f_0$ both detectors are restricted to Jovian‐mass or larger companions, but DECIGO still maintains a clear advantage, underscoring how instrumental sensitivity and mission duration combine with the source’s intrinsic frequency to govern the discovery space for CBPs in the GW band.

\section{Conclusion}\label{sec:conclu}
    This work assesses the feasibility of CBPs around compact binaries from the phase/frequency modulations they imprint on space-borne GW signals. Using a Fisher-matrix framework and a pragmatic detectability criterion based on the fractional uncertainties of the planetary semi-amplitude and period, we mapped the dependence of the minimum detectable planetary mass varies with the planet’s orbital period, the host-binary chirp mass and mass ratio, the source distance, and the carrier GW frequency. We reported forecasts for both B-DECIGO and DECIGO under representative mission durations.
    
    Our main results can be summarized as follows. (i) Sensitivity depends strongly on the planetary period: it improves from very short periods to an intermediate regime where multiple orbits are sampled and the Doppler imprint is strongest, then degrades once the period approaches or exceeds the observing time; a feature near one year reflects degeneracy with the detector’s annual motion. (ii) Larger chirp masses and more equal component masses enhance sensitivity by boosting the carrier SNR and the relative strength of Doppler sidebands. (iii) Increasing distance reduces sensitivity in step with SNR, making massive, nearby binaries prime targets. (iv) For B-DECIGO, favorable Galactic BNS systems reach the super-Earth/Neptune regime under our criterion; DECIGO’s lower noise and longer integration extend the reach toward Earth-mass planets for nearby, high-frequency BNSs and maintain sub-Jovian sensitivity at much greater distances of $\sim 1$ Gpc scale, well beyond the Local Supercluster, even reaching the Abell galaxy clusters.
    
    A note on the DECIGO estimates is warranted. Our DECIGO sensitivity should be regarded as a coarse, trend-level guide rather than a final forecast. Because the assumed observing time (\(T_{\rm obs}\gtrsim4\,\mathrm{years}\)) far exceeds the detector’s one-year heliocentric orbit, we treated the four DECIGO clusters as equivalent and combined their information content by summing identical Fisher matrices, \(\Gamma_{\rm net}\simeq4\,\Gamma_{\rm cluster}\), implying an SNR gain of \(\sqrt{4}\). This simplification neglects the time-dependent constellation geometry and possible inter-interferometer correlations. In addition, the current baseline design of DECIGO and B-DECIGO adopts a three-spacecraft Fabry–Pérot Michelson configuration rather than a LISA-type time-delay interferometry (TDI) scheme \citep{kawamuraCurrentStatusSpace2020}; hence, our analysis adopts a single-Michelson response per cluster. This approximation neglects the detailed orbital motion and inter-interferometer coupling, but remains consistent with the present mission design. A more comprehensive forecast would model the full, time-dependent response of all clusters and channels, which could slightly modify the quantitative sensitivities or parameter-estimation accuracies.

    Several limitations qualify these forecasts. We rely on the linear-signal (Fisher) approximation, which can be optimistic at modest SNRs or near parameter degeneracies (e.g., around one year). We restrict attention to circular, coplanar, single-planet systems, and we do not quantify biases from waveform systematics (e.g., higher-PN terms, spin-induced modulations, tides) or from simplifications in the detector response. Representative source orientations are fixed; we do not marginalize over sky locations, duty cycles, or non-stationary instrumental systematics.
    
    Looking ahead, several extensions can sharpen and stress-test these conclusions: (1) an end-to-end forward model for DECIGO that includes the time-dependent response of all four clusters and their multiple Michelson interferometer outputs, with network-level correlations and realistic duty cycles, replacing the “equivalent-four” approximation; (2) injection–recovery campaigns with Bayesian inference (beyond Fisher) to validate detectability near degeneracies and at low SNR and to quantify selection effects; (3) expanded signal modeling to include eccentric and mutually inclined orbits, multi-planet architectures, and improved compact-binary waveforms, alongside bias assessments from waveform truncation; (4) population-level yield estimates using astrophysically motivated BNS/DWD distributions and mission timelines, together with confusion and environmental noise budgets; and (5) coordinated strategies with electromagnetic techniques (e.g., RV or timing follow-up for nearby candidates) and with other GW bands (e.g., LISA–DECIGO multiband prospects) to break degeneracies and cross-validate candidates. These steps will turn the trend-level picture presented here into a robust, end-to-end forecast and, ultimately, into data-analysis strategies capable of discovering and characterizing CBPs in the decihertz GW window.

\section{Acknowledgements}
    We would like to thank Xianfei Zhang and Ran Gao for their helpful discussions. This work was supported by the National Key Research and Development Program of China Grant Nos. 2023YFC2206702, and 2021YFC2203001; National Natural Science Foundation of China under Grants Nos.11920101003, 12021003, 11633001, 12322301, and 12275021; the Strategic Priority Research Program of the Chinese Academy of Sciences, Grant Nos. XDB2300000 and the Interdiscipline Research Funds of Beijing Normal University.

\bibliography{v2_ref}{}
\bibliographystyle{aasjournalv7}

\end{document}